\documentclass[fleqn,10pt]{wlscirep}
\title{Band bending profile and band offset extraction at semiconductor-metal interfaces}
 
\author[1]{Sergej Schuwalow*}
\author[2]{Niels B. M. Schröter*}
\author[3]{Jan Gukelberger}
\author[4,5]{Candice Thomas}
\author[2]{Vladimir Strocov}
\author[3]{John Gamble}
\author[2]{Alla Chikina}
\author[2]{Marco Caputo}
\author[2]{Jonas Krieger}
\author[4]{Geoffrey C. Gardner}
\author[3]{Matthias Troyer}
\author[2,6,7]{Gabriel Aeppli}
\author[4,5,8]{Michael J. Manfra}
\author[1]{Peter Krogstrup$^{+}$}
\affil[1]{Center for Quantum Devices, Niels Bohr Institute, University of Copenhagen and Microsoft Quantum Materials Lab Copenhagen, Lyngby, Denmark}
\affil[2]{Paul Scherrer Institut, Swiss Light Source, CH-5232 Villigen PSI, Switzerland}
\affil[3]{Microsoft Quantum, One Microsoft Way
Redmond, WA 98052, USA}
\affil[4]{Microsoft Station Q Purdue, Birck Nanotechnology Center, Purdue University, West Lafayette, IN 47907, USA}
\affil[5]{Department of Physics and Astronomy, Purdue University, West Lafayette, IN 47907, USA}
\affil[6]{Physics Department, ETH CH-8093 Zurich, Switzerland}
\affil[7]{Institut de Physique, EPFL CH-1015 Lausanne, Switzerland}
\affil[8]{School of Electrical and Computer Engineering and School of Materials Engineering, Purdue University, West Lafayette, IN 47907, USA}
\affil[*]{these authors contributed equally to this work}
\affil[+]{corresponding author: krogstrup@nbi.dk}
\begin{abstract}
The band alignment of semiconductor-metal interfaces plays a vital role in modern electronics, but remains difficult to predict theoretically and measure experimentally. For interfaces with strong band bending a main difficulty originates from the in-built potentials which lead to broadened and shifted band spectra in spectroscopy measurements. In this work we present a method to resolve the band alignment of buried semiconductor-metal interfaces using core level photoemission spectroscopy and self-consistent electronic structure simulations. As a proof of principle we apply the method to a clean in-situ grown InAs(100)/Al interface, a system with a strong in-built band bending. Due to the high signal-to-noise ratio of the core level spectra the proposed methodology can be used on previously inaccessible semiconductor-metal interfaces and support targeted design of novel hybrid devices and form the foundation for a interface parameter database for specified synthesis processes of semiconductor-metal systems.

\end{abstract}
\begin{document}

\flushbottom
\maketitle
\thispagestyle{empty}
Semiconductor-metal (SM) interfaces play a key role in the functionality of electronic devices. Since the 1960s, attempts were made to obtain reliable predictions for their key characteristic, the band alignment, from charge neutrality points of metal-induced gap states (MIGS) \cite{Hei65, Ter84}, defect levels \cite{All86}, or interface reactions \cite{Fre81}. However, realistic theoretical models typically require detailed information about the interface chemistry that depends on the specific preparation conditions, which makes accurate predictions for the band alignment of a given interface notoriously difficult.\\
An important experimental probe of band alignment is photoelectron spectroscopy (PES), which has the advantage over other techniques - such as electronic transport experiments - that it can directly access the electronic band structure at the SM interface. A common method to extract the band offsets from PES spectra is to either measure the distance of the valence band maximum (VBM) to the Fermi-level (which also gives the conduction band offset by adding the band gap energy), or to measure the binding energy of a core level if its energetic separation to the VBM is known\cite{Kin10,Bri86,Ebe78,Egl84,Him83}. However, comparison between calculations based on the obtained band offset parameters and independent experimental observations, where available, has generally led to poor agreement and suggests systematic errors \cite{Kin07,Kin10}.\\
In terms of band alignment extraction there are two types of interfaces. The simplest case is when the Fermi level at the interface pins in the semiconductor band gap. At low temperatures and in absence of doping this leads to a negligible band bending due to lack of free carriers. If, however, the Fermi level at the interface is pinned outside the gap it gives rise to band bending due to accumulation of carriers (either holes or electrons depending on the sign of the band bending).
\begin{figure*}[p]
\vspace{0.2cm}
\includegraphics[width=0.45\textwidth]{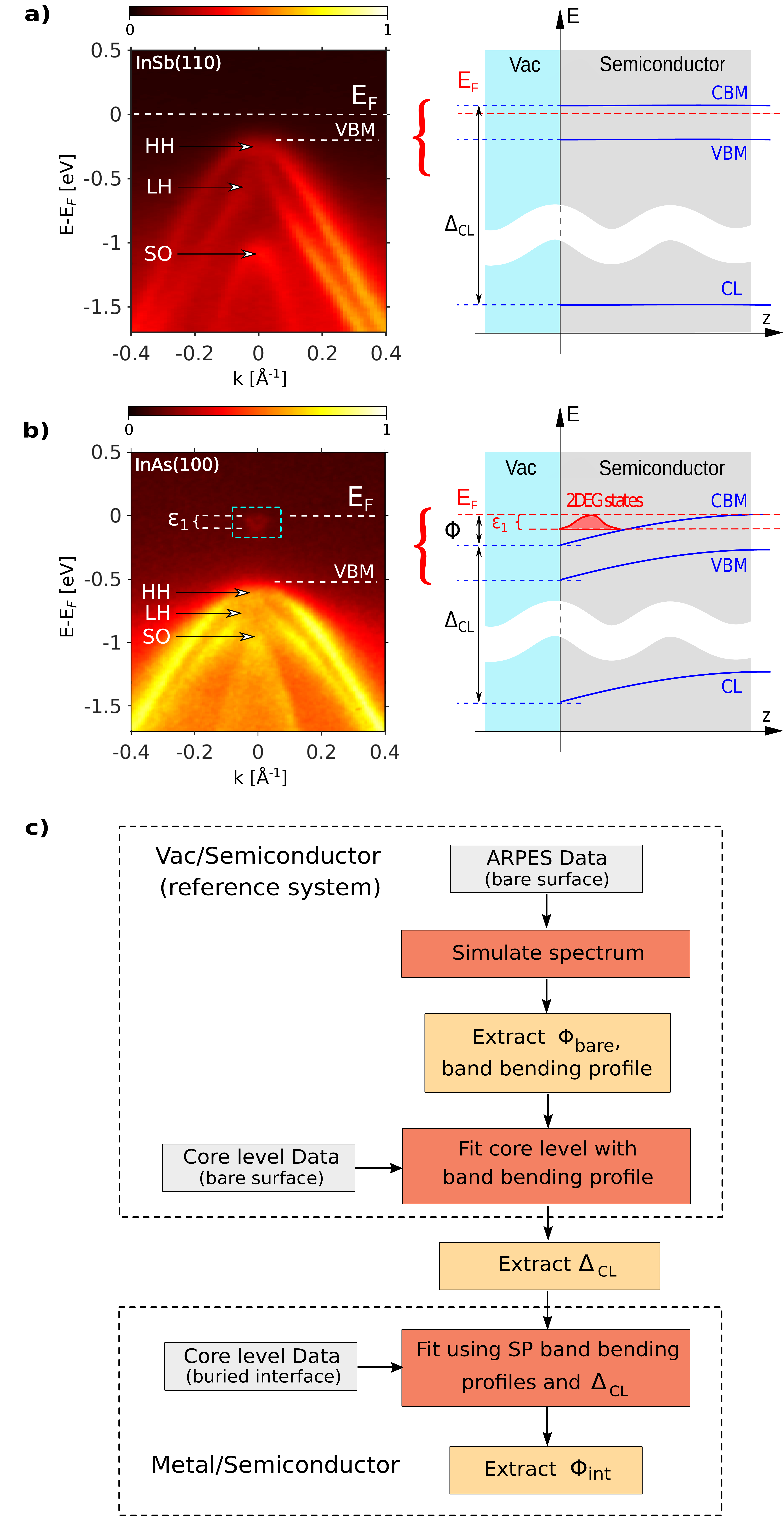}
\vspace{0.2cm}
\caption{a) ARPES band structure (h$\nu$ = 707 eV) and band alignment diagram for a system with surface Fermi level pinning within the gap, here exemplified on the InSb(110) system b) Same as a) but for InAs(100), a system with the Fermi level pinned above the CBM and strong band bending (h$\nu$ = 405 eV). $\epsilon_1$ denotes the visible offset of the first QW subband relative to the Fermi level $\textrm{E}_\textrm{F}$, with the usual abbreviations for valence band maximum (VBM), heavy-hole (HH), light-hole (LH) and split-off (SO) bands. Emergent 2DEG states are highlighted. c) Flow diagram of the band offset determination procedure for buried interfaces with band bending.  }
\label{fig1}
\end{figure*}
Fig.\ref{fig1} presents the two types of systems characterized by soft X-ray angle-resolved photoemission spectroscopy (SX-ARPES). In Fig.\ref{fig1}a we show the that the Fermi level of the InSb(110) surface is pinned in the gap as illustrated in the corresponding diagram. The valence band structure consists of the heavy-hole (HH), light-hole (LH) and split-off bands (SO), with the valence band maximum (VBM) visible at $\sim0.2$eV below the Fermi level (for details see \hyperref[APP-Method]{Methodology}). In absence of an accumulation layer the bands have a flat depth profile on the length scale relevant for device physics (without a significant doping of the semiconductor substrate). This flat profile was verified by core level spectroscopy which independently showed constant core level binding energy as a function of depth.\\
In contrast to the pristine InSb surface, the pristine InAs(100) surface, as shown in Fig.\ref{fig1}b, is known to exhibit a downwards band bending towards the surface\cite{Pip06,Kin10}. The behavior of this system is more complicated in the sense that the Fermi level is pinned in the conduction band and the resulting quantum well contains a 2-dimensional electron gas (2DEG) confined to the surface. All states in the system including the core levels (CLs) are uniformly affected by this confining potential as an in-built field. The quantum well states are visible just below the Fermi level with subband energies $\epsilon_{i}$ at the $\Gamma$ point (at k=0) where \textit{i} is the subband number. $\Phi$ denotes the conduction band offset. Note that the bottom of the lowest quantum well subband $\epsilon_{1}$ is \textit{not} equal to the band offset $\Phi$ because of the additional energy increase due to confinement.\\
The usual approach to obtain the band offset is based on the determination of the VBM using the intersection of the linear fit of the background in the band gap\cite{Cha04} with the extrapolation of the valence band leading edge. However, even in the flat band case accurate determination of the VBM is not straightforward: the background in the gap can be nonlinear due to band broadening, especially in the case of narrow gap semiconductors. Measurement of a core level as a substitute for the valence band requires the a-priori knowledge of the energy separation between the VBM and the core level.\\
Additional complications arises in systems with band bending. Typical PES measurements with electron energies of 10-1000 eV accumulate 95\% of their detected photo-electrons from an extended subsurface region with a depth of $\sim$1.5-6 nm. Thus the confining potential of the surface affects all photoemission spectra, both valence band and core level through asymmetric broadening and shifting of peak positions, especially in systems with a narrow quantum well and strong band bending at the interface \cite{Kin07,Bai08}. Deconvolution methods for these band-bending modulated PES spectra were developed in the past\cite{Kak08, Vio13, Tan11, Wip16, Min12, Du18} using analytical potential shapes or interface models. However, the band bending effect can be challenging to disentangle from other sources of line broadening, such as lifetime effects, instrumental resolution or chemical shifts \cite{Sok58}. The small effective material volume from which information can be obtained makes this deconvolution procedure an ill-posed mathematical problem.
As a consequence, Schr\"odinger-Poisson calculations used to simulate accumulation layers based on measurements on CdO, InAs, and InN resulted in an underestimation of the quantum well carrier density and subband occupation number and led to errors in the conduction band offset $\Phi$ of $\sim$100 \% \cite{Kin07,Kin10}. 
These issues are amplified for buried interfaces by the attenuation due to the overlayer. Depending on the intensity and energy of the available light source and the specifics of the band structure, conduction and valence band measurements quickly become unfeasible with increasing overlayer thickness. Core levels with their strong spectroscopic signatures remain the only accessible features for buried interfaces.\\
The approach to band bending profile analysis of interfaces presented in this work combines angle-resolved photoelectron spectroscopy, photon energy dependent core-level spectroscopy and self-consistent electronic structure calculations. The method avoids many systematic errors made in the analysis of PES spectra from heterostructures and allows investigation of buried interfaces previously out of reach.
The approach is outlined in Fig.\ref{fig1}c and consists of two parts; 1) determination of an accurate value of the characteristic bulk energy difference between a core level and the conduction band of the reference semiconductor $\Delta_{\textrm{CL}}$, and 2) core level measurements on the buried interface system of interest (based on the same semiconductor). It is important to note that the energy difference $\Delta_{\textrm{CL}}$ is independent of the overall potential and therefore a constant property of the bulk material.\\
To determine $\Delta_{\textrm{CL}}$ we first use angle-resolved photoelectron spectroscopy (ARPES) to obtain momentum resolved dispersion of quantum well subbands. Bare semiconductor surfaces with an accumulation layer (such as the InAs(100) surface\cite{Ols96PRL}) are suitable for this step as they give the highest signal-to-noise ratio for the emission from the conduction band. In case the bare surface does not have an accumulation layer, it may be induced intentionally by changing the surface chemistry as shown in ref[\cite{Bet96}] in the case of InSb. Knowing the material dimensions of the stack and the energy levels of the quantum well states, we can use a self-consistent electronic structure model to obtain the conduction band offset $\Phi_{bare}$ (see the definition sketched in Fig.\ref{fig1}b) and the band bending profile. After the determination of the conduction band profile a fitting procedure is applied to a complementary set of core level spectra. The difference between conductance band and the chosen core level gives the characteristic energy difference $\Delta_{\textrm{CL}}$ (compare panel b)). \\
In the second part of the procedure, a set of PES spectra from the \textit{same} CL is acquired on the buried interface. We fit this dataset with a band bending profile from a set of possible self-consistent solutions of the electronic structure model for the given heterostructure geometry, shifted by the previously determined $\Delta_{\textrm{CL}}$. This effectively restricts the fit to the space of physically consistent profiles which are, via the model fit procedure, themselves functions of the buried interface offset $\Phi_{int}$. In the following we show that this approach addresses the issues discussed above and makes the core level fitting procedure reliable for the purpose of band offset determination. In this work we use Schr\"odinger-Poisson (S-P) simulations\cite{Tre97}, however, alternative approaches such as the $k\cdot p$ formalism (or many-body methods) can be used depending on the type of system and level of detail.\\
\begin{figure*}[p]
\vspace{0.2cm}
\includegraphics[width=0.3\textwidth]{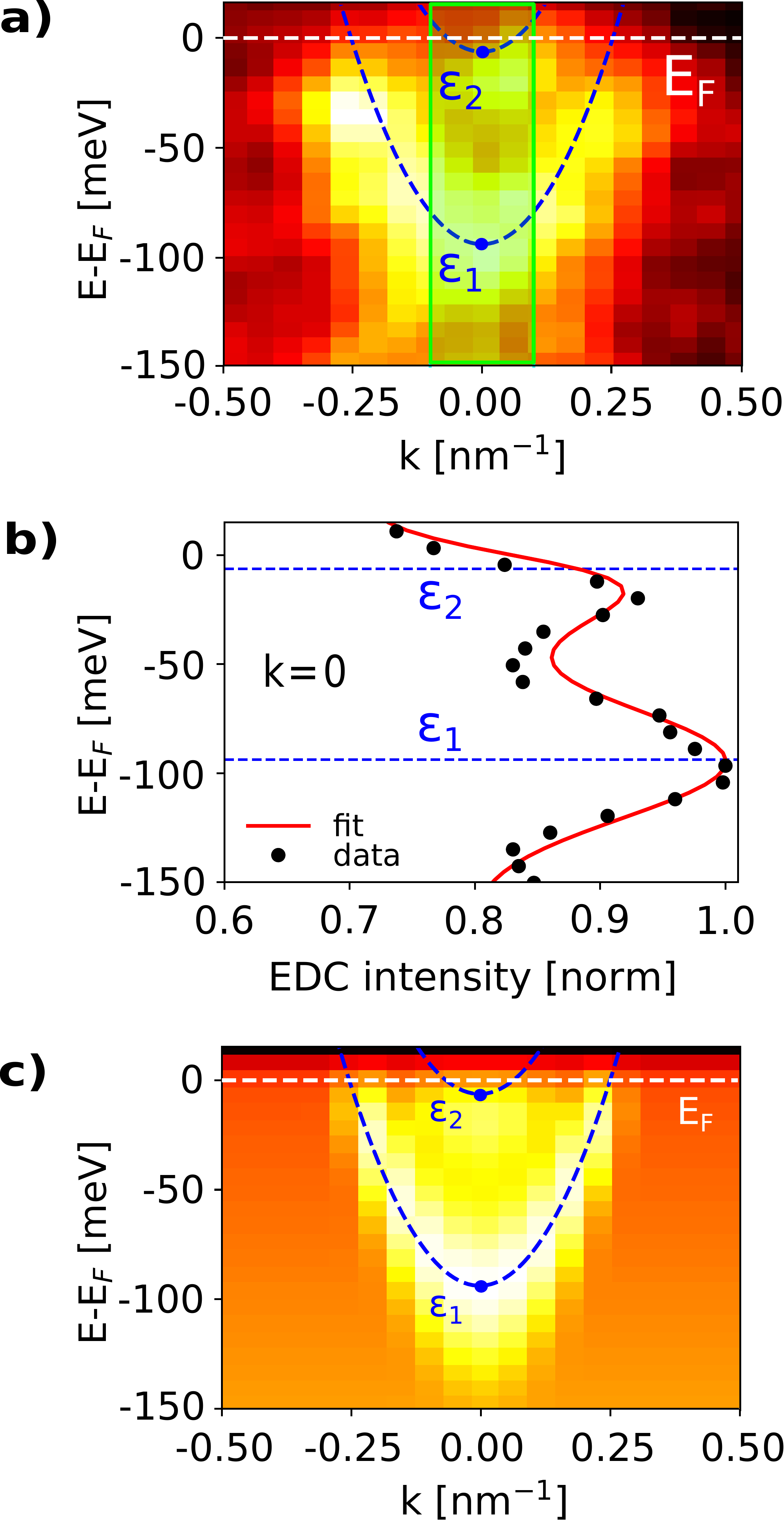}
\vspace{0.2cm}
\caption{a) Zoom-in of the QW states highlighted in Fig.\ref{fig2}b. The blue dashed lines show the dispersions of the lowest two QW states in a Schr\"odinger-Poisson model fit to the data, which yields the band offset $\phi = 0.019(3)$. The fit area is shown as a green rectangle.
b) Energy distribution curve (EDC) for the experimental spectrum shown in a) at k=0. Black dots represent the data. The red solid line shows the SP fit, with the EDC profile described by two Gaussian functions with linear background. Dashed blue lines indicate the energy levels $\epsilon_{1,2}$ obtained from the SP fit. c) Simulated ARPES spectrum based on the SP fit shown in a),b) 
}
\label{fig2}
\end{figure*}
In the following we apply the method to the InAs(100)/Al interface which has recently gained interest for its potential in topological quantum computing \cite{Nay08,Min16,Lut18,Sar15,Nic17}. Detailed knowledge of the electronic band alignment at the interface is essential for the understanding of superconductor-semiconductor hybridization in these systems \cite{Mik18}. We demonstrate that the energies and occupation numbers of the quantum well subbands at the SM interface can be extracted with an error of $\pm 10$\% for the band offset. The first step of the procedure is the acquisition of the $\Delta_{\textrm{CL}}$ value using the pristine InAs(100) surface (Fig.\ref{fig1}b). This is done by first fitting the in-plane QW state dispersion from the conduction band (around the $\Gamma$ point) shown in in Fig.\ref{fig2}a (compare dashed rectangle in Fig.\ref{fig1}b). The two dashed blue lines in Fig.\ref{fig2}a show the Schr{\"o}dinger-Poisson fit of the data within the area highlighted in light green (see \hyperref[APP-SP]{Supplementary A} for details). The corresponding energy distribution curve (EDC) for the $k=0$ data bin is shown in Fig.\ref{fig2}b. From this procedure we obtain a band offset of $\Phi_{InAs(100)}=0.19(3)\textrm{eV}$, with the subband energies \textrm{$\epsilon$}$_{1}=-0.09\textrm{eV}$ and \textrm{$\epsilon$}$_{2}=-0.006\textrm{eV}$, respectively. The simulated ARPES spectrum resulting from the fit procedure described in \hyperref[APP-SP]{Supplementary A} is shown in Fig.\ref{fig2}d. \\
The corresponding core level data is recorded at the same position as the angle resolved spectra using a photon energy range of 350 - 1050 eV (Fig.\ref{fig3}a). The ratio of bulk to surface contribution varies with photon energy because of the photon energy dependent penetration depth. This allows the core level fitting procedure to separate the surface and bulk contribution as well as capture the trend of the band bending. Any well-defined core line can be chosen for this procedure depending on the material - here we use the In4d core level. Fig.\ref{fig3}a shows the core level data set for the bare InAs. The shape of the In4d level in InAs has been subject to discussions in literature in the past\cite{Ols96,Ols96PRL}. It is known that the spectral line consists of two distinct components, each exhibiting a two-peak shape caused by the spin-orbit interaction; a main component originating from the bulk and a smaller contribution stemming from the surface layer of the material. The surface component exhibits a shift in energy due to the different local environment and bond formation of the surface. The bulk-like component contains information about the band bending profile. The core level fitting procedure itself is described in \hyperref[APP-Model]{Supplementary C}. The core level profiles in the entire energy range are simultaneously fitted using the potential profile obtained from the Schr\"odinger-Poisson simulation of the QW states and a rigid energy shift. This represents the in-built field simultaneously affecting all states of the spectrum and gives the characteristic core level energy $\Delta_{CL} = {-}17.22(3)\textrm{eV}$ for the InAs In4d core level.\\ 
Having obtained the value for $\Delta_{\textrm{CL,In4d}}$ for InAs we can study all types of buried interfaces involving InAs. Here we investigate the InAs(100)/Al interface. Fig.\ref{fig3}b shows the change in the In4d line shape upon Al deposition. The most immediate effect of the Al deposition is the emergence of a distinct 3-peak structure, with the additional component now located at lower binding energies than the bulk feature. Analysis of the relative intensity modulation between the surface and bulk components (see \hyperref[APP-In]{Supplementary B}) demonstrates that this low-energy component can be attributed to traces of metallic In located \textit{on top} of the deposited Al layer. This behavior is consistent with the relative weakness of In-Al bonds, and the system minimizing energy by removing In from the surface in favor of As-Al bond formation. InAl, to our knowledge, is not known to form any stochiometric compound, and the strongly different size factors of the involved atoms seem to result in a migration of excess In atoms to the upper layers of the deposited Al thin film, similar to what has been previously observed for InSb/Al \cite{Spo88,Bos87}.\\
\begin{figure*}[p]
\vspace{0.2cm}
\includegraphics[width=0.45\textwidth]{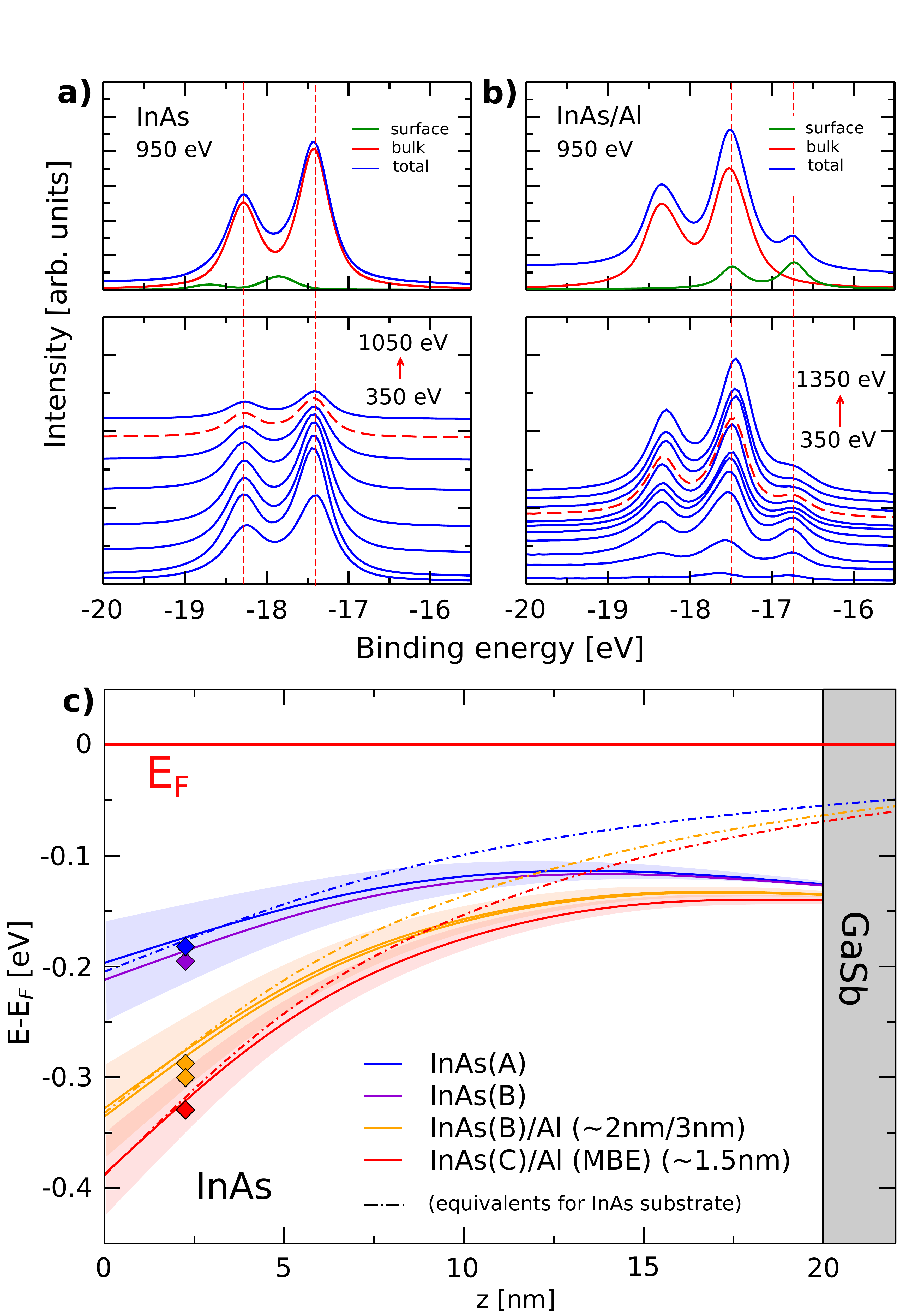}
\vspace{0.2cm}
\caption{a) In4d core level data for photon energies between 350 eV and 1050 eV for the pristine InAs surface (Sample A). The top panel shows the decomposition of the In4d level into the surface and bulk components. The bottom panel shows the raw data, shifted vertically for convenience. The dataset shown in the top panel is highlighted with a dashed red line. b) Same, but for InAs with $\sim$2 nm of Al deposited inside the ARPES chamber (Sample B), shown for photon energies between 350 eV and 1350 eV. c) Band bending potentials for the InAs(100)/Vac and InAs(100)/Al interfaces. Same-sample results are color-coded, with the confidence intervals shown in the background. Full lines correspond to the 20nm InAs slab used in the experimental setup. Dashed lines show equivalent calculations assuming semi-infinite InAs slab.
}
\label{fig3}
\end{figure*}
To gain more understanding of the InAs/Al system we compare three different samples; a pristine InAs (Sample A) surface analyzed in Fig.\ref{fig2}, a separate sample where Al layers of varying thickness are deposited onto the freshly decapped InAs substrate directly inside the ARPES preparation chamber (Sample B, clean surface, and with $\sim2$/$\sim3$ nm Al) and an in-situ MBE-grown InAs/Al layer (Sample C, $\sim1.5$ nm Al). Since we know $\Delta_{\textrm{CL,In4d}}$ for the material in question and use physically self-consistent profiles obtained by the SP approach the fitting procedure becomes a well-defined mathematical problem with a single parameter, the interface band offset $\Phi_{int}$.\\
Both the reference InAs fit and the results of the core level fitting procedure for the subsequent samples are shown in Fig.\ref{fig3}c. The pristine InAs surface demonstrates a downwards band-bending of approx. 0.2 eV, in line with the band bending strength suggested (but not observed in the CL data) by King\cite{Kin10} \textit{et al}. For the buried InAs/Al interface, the model clearly demonstrates a further increase of the band offset by approx. 0.1-0.2 eV. Note that the in-situ MBE-deposited Al film (C) which is known \cite{Kro15} to produce very clean epitaxial interfaces, shows a comparatively stronger effect of Al deposition than the film deposited inside the ARPES chamber (B), which we speculate to be a consequence of higher interface quality. We do not detect a difference in the measured band bending profile between the 2 nm and 3 nm Al deposition in the ARPES chamber outside of our error margins. The results are summarized in Tab.\ref{tab:Fit_results}.\\ 
For context, the color-coded squares in Fig.\ref{fig3}c show the directly observed energies of the core level maxima without the band bending model decomposition. The dashed lines show the calculated band bending profiles for a semi-infinite InAs slab instead of a 20 nm thick substrate used in the experiment. Note that the band offset is a property of the interface and thus unaffected by the details of sample geometry. \\
 \begin{table}[b]
 \centering
 \begin{tabular}{l|l|l}
Material & Sample & $\Phi$[eV]\\
 \hline
 InAs & A    & -0.19(3)\\
 \hline
 InAs & B    & -0.21(3) \\
 \hline
 InAs / Al ($\sim2$nm/$\sim3$nm) & B    & -0.33(3) \\
 \hline
 InAs / Al (MBE) & C  & -0.39(4) \\
 \end{tabular}
 \caption{
 Overview of band offset parameters $\Phi$ for the investigated heterostructures.}
 \label{tab:Fit_results}
 \end{table}
To validate the procedure we perform a direct SX-ARPES measurement of a SM interface (Sample C) with a metal layer thickness that is thin enough to still allow access to the quantum well states (Fig.\ref{fig4}a). To compensate for the attenuation by the Al layer the measurement is performed at the higher photon energy, here $hv=1045$ eV, which results in an increased electron mean free path (at the cost of reduced energy resolution). In Fig.\ref{fig4}b we zoom in on the conduction band QW states. The overlaid blue dashed lines are the QW states obtained independently by the core level fitting and self-consistent SP approach. A comparison between these and a direct fit of the EDC, for the k=0 bin, is displayed in Fig.\ref{fig4}c. The core level fitting procedure predicts the energy levels of the first and second QW state as \textrm{$\epsilon$}$_{1,CL}=-0.17\textrm{eV}$ and \textrm{$\epsilon$}$_{2,CL}=-0.07\textrm{eV}$. This is in good agreement with values obtained from fitting the direct measured states (\textrm{$\epsilon$}$_{1,direct}=-0.16\textrm{eV}$ and \textrm{$\epsilon$}$_{2_,direct}=-0.06\textrm{eV}$), even though the resolution of the quantum well states is low due to the attenuation by the metal layer. The obtained band offsets of $\Phi_{int,CL}=-0.39(4)\textrm{eV}$, and $\Phi_{int,direct}=-0.35(5)\textrm{eV}$, respectively, confirm that the accuracy of the method is comparable with the accuracy obtained by direct fitting of the CB states (when accessible). We emphasize that this result is obtained using only a sequence of core level measurements on the buried interface system. Given high core level line intensity, we expect that an accurate determination of band offsets in buried interfaces under up to 6-8 nm of metallic overlayers should be possible in the soft-X-ray photon energy range.\\
\begin{figure*}[p]
\vspace{0.2cm}
\includegraphics[width=0.3\textwidth]{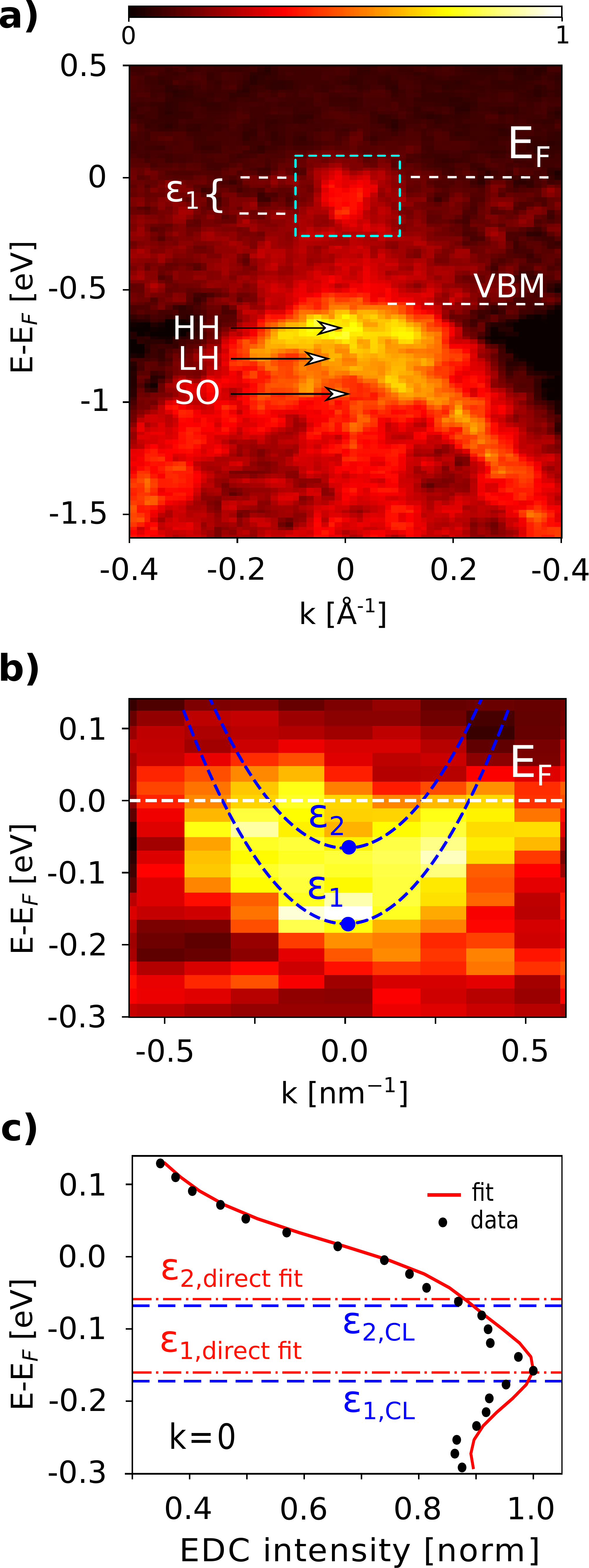}
\vspace{0.2cm}
\caption{Verification by comparison with direct measurement. a) Background-subtracted SX-ARPES spectrum of the buried interfacial electronic structure of MBE grown InAs/Al($\sim$1.5nm) heterostructure (Fig.\ref{fig3}c, Sample C) at hv = 1045 eV (compare with Fig.\ref{fig2}a) b) A zoom in on the QW states of the InAs/Al system, overlaid with states calculated independently with input from \textit{only} the core level fitting procedure, c) EDC for the InAs/Al experimental spectrum at k=0. The red full/dashed lines shows a direct SP fit and the energy levels obtained therefrom (compare with Fig.\ref{fig2}c), blue dashed lines show the energy levels obtained from the core level fitting procedure.
}
\label{fig4}
\end{figure*}
There are two central points which allow the procedure presented in this work to function at this level of accuracy; the choice of a good reference system for the determination of $\Delta_{\textrm{CL}}$ and the use of self-consistent SP potentials in conjunction with $\Delta_{\textrm{CL}}$ in the core level fitting. We suggest the conduction band states of an electron accumulation layer as the reference system because of their clear signature suitable for fitting and their well-defined relation to the confining potential. Any experimentally obtained quantity that can uniquely determine an SP solution on \textit{any one} surface based on the reference substrate can also fulfill the same purpose. However, accuracy in the treatment of the reference system is important as all uncertainties will propagate through the analysis.\\
The use of self-consistent SP potentials throughout the fitting procedure guarantees a physically sound relationship between band offset $\Phi$ and the band bending profile that enters the core level model (\hyperref[APP-Model]{Supplementary C}). Because of this the core line shape and the core level binding energy are no longer independent fitting parameters up to a fixed energy difference which is a material property of the bulk system ($\Delta_{\textrm{CL}}$). This presents a very strong fitting constraint. If $\Delta_{\textrm{CL}}$ is known the offset $\Phi$ can be determined to a very high accuracy, and vice versa.\\
In summary, the method demonstrated in this work can serve as a general approach for extracting reliable values for a) $\Delta_{\textrm{CL}}$, characteristic bulk parameter of the semiconductor, and b) band offsets $\Phi$, key parameters of SM interfaces.\\ 
The band offset of SM interfaces is the key parameter for engineering SM interface in most electronic devices. In the case of semiconductor/superconductor materials like InAs/Al, the band offset is likewise a critical parameter for hybridization at the interface heterostructures. While the direct measurements of the \textit{interface} QWS in this work are used as a verification of the fitting result (Fig.\ref{fig4}), the procedure is not a requirement. The core level offset $\Delta_{\textrm{CL}}$ does require both core level and ARPES (or equivalent) measurements from the pristine substrate (or an alternative reference system), but once the reference system has been measured to the required degree of accuracy the transition from the bare band bending profiles to those of buried interfaces can be done with core level spectra only. The core lines typically provide orders of magnitude higher photoemission signal as compared to conduction or valence band photoemission. This drastically reduce the associated acquisition times and/or lowering requirements as to heterostructure thickness and surface quality. It further removes the need for angular resolution, thus increasing accessibility. We believe that the combined advantages of this approach can strongly contribute to fast development of engineering novel material combinations and targeted heterostructure interface design in the future.

\bibliography{main}

\section{Acknowledgements}
We would like to thank Roman Lutchyn, Georg Winkler, Jordi Arbiol, Scott Chambers and K. Flensberg  for fruitful discussions.
This work was supported by the Microsoft Station  Q  Program  and  by  the  European  Research Commission, project HEMs-DAM No. 716655, starting grant under Horizon 2020. A.C. and M.C. were supported by the Swiss National Science Foundation under grant No. 200021\_165529, and J.A.K. under grant No. 200021\_165910.

--
\section{Additional information}

\subsection{Methodology}\label{APP-Method}
The planar InAs-based materials were grown by molecular beam epitaxy on GaSb (100) substrates in a Veeco Gen 930 using ultra-high purity techniques and methods as described in \cite{Gar16,Tho18}. The structures are composed of a GaSb buffer grown at 500$^{\circ}$C followed by a 20 nm-thick InAs layer grown at 480$^{\circ}$C, as measured by absorption of blackbody radiation. The transition between these two materials has been made using a shutter sequence developed in \cite{Tut90}. The growth of InAs was performed with an As-to-In beam equivalent pressure ratio slightly larger than 1 to prevent the formation of void defects associated with As etching of the GaSb layer. Under these conditions, flat InAs surface morphology has been obtained with a roughness of the order of a monolayer.\\
To facilitate the observation of the quantum well states, the first d=15 nm of the InAs layers were doped with Si atoms at a density n$_{\textrm{Si,3D}} = 2.2\times10^{18}\textrm{cm}^{-3}$. After the growth, the samples were covered with an amorphous arsenic cap layer deposited at 0$^{\circ}$C to protect the InAs surface during transport in nitrogen atmosphere from the MBE-system in Purdue to the SX-ARPES system at the ADRESS beamline of the Swiss Light Source. The InAs samples were decapped by annealing at $\sim350^{\circ}$ C in the ultra-high vacuum system at ADRESS. \\
During the ARPES in-situ deposition steps described in the second part of the paper Al layers with different thicknesses were deposited at a temperature of T$\sim$15 K on top of the pristine InAs(100) by employing a shadow mask technique. The layer thickness was checked post deposition as a part of the fit described in \hyperref[APP-Model]{Supplementary C}.\\
The InSb(110) surface shown in Fig.\ref{fig1}a was prepared by cleaving inside the ARPES preparation chamber.

\subsection{Supplementary A: Schr\"odinger-Poisson fit of the conduction band spectrum}\label{APP-SP}
 
Electrostatic potential profile and subband energies are obtained from a self-consistent Schrödinger-Poisson solver modeling the conduction, heavy-hole and light-hole bands \cite{Tre97}. We include the InAs layer and a 100 nm thick slice of the GaSb substrate in the calculation, with space discretization along the growth direction $\Delta z \leq 0.2$ nm. The band offset $\Phi$ enters the Poisson equation as a Dirichlet boundary condition at the semiconductor surface, whereas a Neumann condition is imposed at the bottom of the system. Band structure parameters, including the conduction band effective mass $m_{*}^{\text{InAs}} = 0.026$ and the InAs-GaSb valence band offset of 0.56 eV, are taken from \cite{Vur01}.
 
Given a S-P solution, a predicted ARPES intensity profile is calculated by convolving the native signal
\footnote{We neglect higher subbands crossing the band offset only for band offsets $\Phi \gtrsim 0.6$ eV as well as Lorentzian broadening.}
$I_{\text{int}}(E, k) = \left[ \sum_{i=1,2} a_i \exp(-\frac{[E - \epsilon_i(k)]^2}{\sigma_{\text{int}}^2}) \right] \theta(-E) + a_{\text{bg}} E + b_{\text{bg}}$
with the instrument resolution (Gaussian broadening in energy and momentum space according to the beamline resolution calculator). $\theta(-E)$ is the Fermi-Dirac step. Since the subband dispersions $\epsilon_{1,2}(k) = \epsilon_{1,2} + \frac{\hbar^2 k^2}{2 m_*^{\text{InAs}}}$ depend on the band offset $\Phi$ via the S-P solution, we obtain $\Phi$ along with all auxiliary parameters (subband intensities $a_{1,2}$, native linewidth $\sigma_{\text{int}}$, and linear background $a_{\text{bg}}$, $b_{\text{bg}}$) from a single least-squares fit of the predicted to the measured intensity profile.

\subsection{Supplementary B: Origin of the shifted In4d core line component after Al deposition}\label{APP-In}

\begin{figure*}[p]
\vspace{0.2cm}
\includegraphics[width=0.5\textwidth]{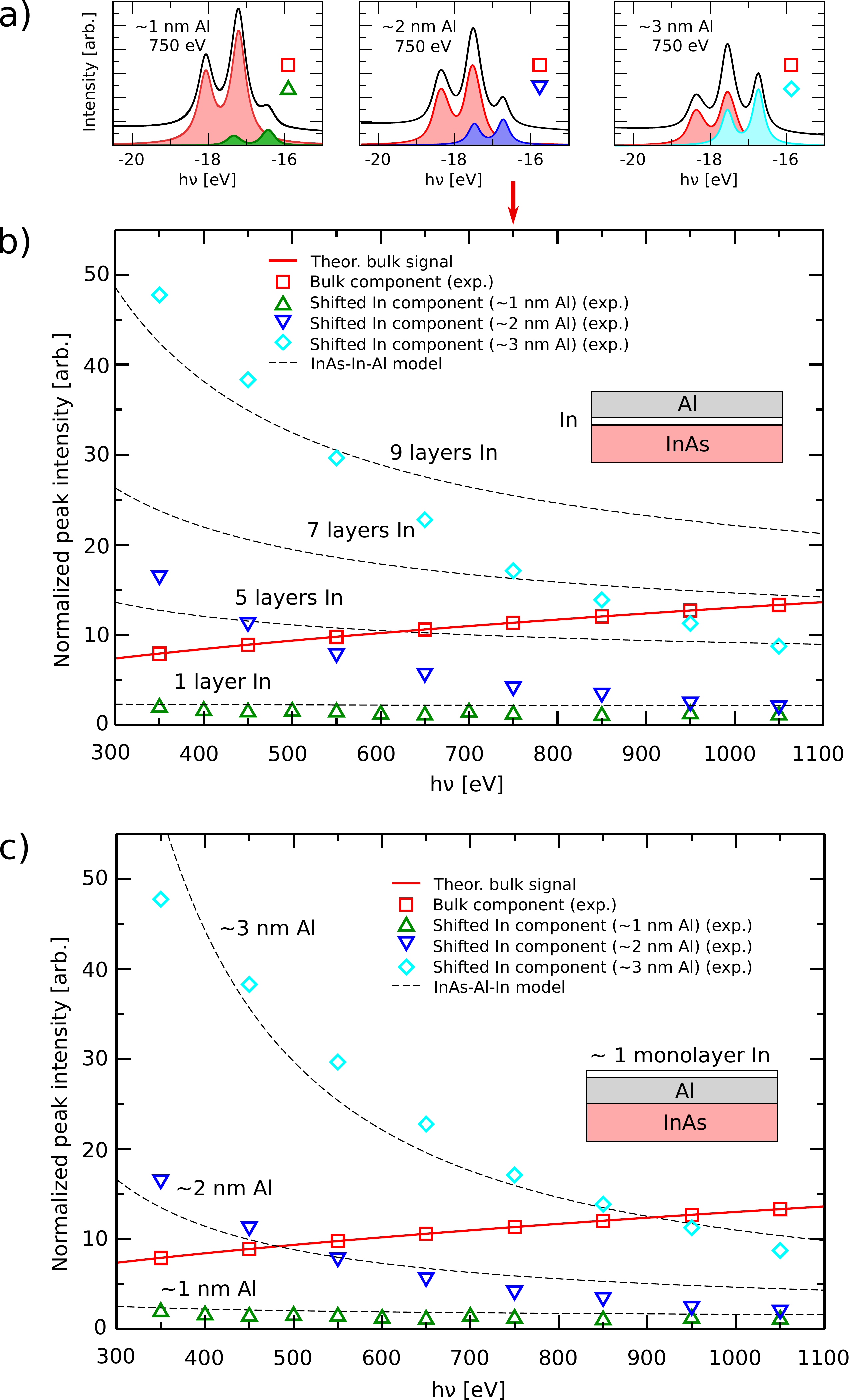}
\vspace{0.2cm}
\caption{a) Phenomenological change in the In4d core level structure for Al deposition thickness of ~1, ~2 and ~3 nm, here for h$\nu=750$ eV. Decomposition in bulk-like and shifted components is shown. b) Normalized intensity of the shifted In4d component vs. theoretical model under the assumption that the signal originates from indium at the InAs/Al interface (InAs-In-Al model). Dashed lines show theoretical intensities from different amounts of In at the interface. No reasonable agreement between model and experiment can be reached, even for unrealistic amounts on In. b) Same, but under the assumption that the shifted In core line originates from the top of the Al layer. Good agreement can be reached between theoretical intensities and experimental observation for a total amount of $\sim$ 1 monolayer of In migrating to the top of the Al layer during deposition.  
}
\label{fig_Supp_A}
\end{figure*}

Deposition of Al on InAs(100) gives rise to an increasingly distinct 3-peak structure of the In4d core level, as shown in Fig.\ref{fig_Supp_A}a for $h\nu=$750 eV and 1/2/3nm Al, respectively. The origin of the shifted core line can be deduced by analyzing the signal for different energies and Al layer thickness.\\
The main core level feature (red) is understood to originate from InAs proper, and thus its total intensity for a given photon energy $h\nu$ is expected to be well described by  $\int_{surface}^{\infty}~I_{0}\exp{(-z\\/\lambda[h\nu])}~dz$, with $I_{0}$ some unit intensity and $\lambda$ the inelastic mean free path of electrons in InAs. To avoid discussion of superimposed intensity variations due to changes of the photon flux with different monochromator settings and the possible dependence of photoemission cross-section matrix elements on energy we renormalize the data to reproduce this behavior in the bulk component and compare the phenomenological behavior of the core-level sub feature with model predictions.\\
Figs.\ref{fig_Supp_A} b), c) show two different models for the intensities of the individual components of the In4d level as a function of photon energy $h\nu$. The symbols correspond to core line components shown in panel a). No reasonable agreement between experimental data and the model can be found under the assumption that the shifted In component originates from the InAs surface (Fig.\ref{fig_Supp_A}b). On the contrary, the  data is readily explained by a total of $\sim$1 monolayer of In being removed from InAs and migrating to the top of the Al layer during Al deposition.

\subsection{Supplementary C: Core level fitting procedure for extraction of band bending profiles}\label{APP-Model}
To determine the band bending profiles shown in Fig.\ref{fig3}c we simultaneously fit the core level shape of a set of spectra in the 350-1350 eV range. The surface component for each photon energy $h\nu$ is represented by
\begin{equation}
S = \int_{0}^{d_{surface}}I_{0}(h\nu)\times\exp{(-z/\lambda)}\times\left[V_{1}(\alpha, \gamma, E-E_s) + V_{2}(\alpha, \gamma, E-E_s+SO)\right]~dz    
\end{equation}
whereas the the bulk component is represented by
\begin{equation}
B = \int_{d_{surface}}^{\infty}I_{0}(h\nu)\times\exp{(-z/\lambda)}\times\left[V_{1}(\alpha, \gamma, E+\Delta_{CL}+\phi(z)) + V_{2}(\alpha, \gamma, E+\Delta_{CL}+\phi(z)+SO)\right]~dz    
\end{equation}
and can be further attenuated by the presence of Al (compare \hyperref[APP-In]{Supplementary B}). Here, V is a Voigt profile with $\alpha$ and $\gamma$ being the FWHM of the Gaussian and Lorentzian components, respectively, SO is the spin-orbit splitting and $E_s$ is the energy of the In4d surface component. $\phi(z)$ is the band bending potential, predicted by the SP model for a given band offset $\Phi$.\\
The inelastic mean free path $\lambda$ is taken from the publications \cite{Tan91, Tan91-2, Tan93} where we use the fits to optical data, if available, and values given by the TPP-2M formula otherwise. The prefactor $I_{0}$ is dependent on energy and absorbs variations in flux, detector sensitivity and photoemission cross-section.

\end{document}